\documentstyle[12pt]{article}
\begin{document}
\begin{titlepage}
\thispagestyle{empty}

\vspace*{2cm}

\begin{center}
{\large \bf A model of short-range correlations in the charge
response }

\vspace{1.5cm}
{\large J. E. Amaro and A. M. Lallena}

\vspace{.3cm}
{Departamento de F\'{\i}sica
Moderna, Universidad de Granada, \\
E-18071 Granada, Spain}

\vspace{1.5cm}
{\large G. Co'}

\vspace{.3cm}
{Dipartimento di Fisica, Universit\`a di Lecce \\  and 
I.N.F.N. sezione di Lecce, I-73100 Lecce, Italy}

\vspace{1.5cm}
{\large A. Fabrocini}

\vspace{.3cm}
{Dipartimento di Fisica, Universit\`a di Pisa \\  and 
I.N.F.N. sezione di Pisa, I-73100 Pisa, Italy}

\end{center}

\vspace{2cm}
\begin{abstract}
The validity of a model treating the short-range correlations up to
the first order is studied by calculating the charge response of an
infinite system and comparing the obtained results with those of a
Fermi Hypernetted Chain calculation.
\end{abstract}

\vskip 1.cm
PACS number(s): 21.60.-n, 24.10.Cn, 25.30.Fj
\end{titlepage}
\setcounter{page}{1}

The interest in the study of Short 
Range  Correlations  (SRC) in nuclear systems has increased in these last
few years. In this field, the experimental activity has been concentrated
in the search for observables  allowing a clean identification of SRC 
effects \cite{ben90,tne95}. 
From the theoretical point of view there has been a development of 
calculations which explicitly consider SRC. 

The theoretical situation is quite satisfactory for the few--body systems
where Faddeev \cite{che86}, Correlated Hyperspherical Harmonics Expansion
\cite{kie93} and Green Function Monte Carlo \cite{car88} theories solve
exactly the  Schr\"odinger equation. This last technique has been recently
applied to investigate light nuclei up to $A=7$ \cite{pud97}.
Unfortunately the straightforward
application of these theories to the study of medium and 
heavy nuclei is not yet technically feasible, 
in spite of the rapid progress of the 
computer technology.

The other satisfactory situation, from the theoretical point of view,
regards the opposite side of the isotope table: the infinite nuclear
systems such as neutron and nuclear matter. For the study of these
systems, perturbations techniques have been developed such as Brueckner
Bethe Goldstone \cite{bal91} or Correlated Basis Function theories (CBF)
\cite{wir88}. These approaches do not provide an exact solution of the
Schr\"odinger equation, but they can reproduce rather well
the empirical properties of nuclear matter because
they sum complete sets of terms of their perturbation expansions.

The application of CBF to the description of the ground state of
finite nuclear systems, has been recently carried out
\cite{ari96} using various levels of Fermi Hypernetted Chain (FHNC) 
approximations. 
The results are promising and it is conceivable that CBF theories may 
be applied to the description of excited states in the future.
For the time being, these properties have to be studied by using 
simpler models.

The major part of the finite models
developed up to now to describe finite nuclear systems
treat SCR only at the lowest orders in the correlation. 
The main field of application has been the
investigation of nuclear ground state properties \cite{kha68}. Recently,
nuclear models dealing with SRC 
have been implemented to study the electromagnetic
two--nucleon emission \cite{giu91,ryc95}.  Given the wide use of these
models and their relevance for predictions and comparisons with two-nucleon
emission data, we believe that a test of their validity is
necessary.

A model treating SRC up to the first order in the correlation, 
has been developed in Ref.~\cite{co95} to study charge density
and momentum distribution of some doubly closed shell nuclei. 
The model was able to reproduce rather well the finite nuclei FHNC
results of Ref.~\cite{ari96} and 
it has been extended to describe the two-nucleon  emission produced 
by the
same electromagnetic operator, the charge operator, for inclusive electron  
scattering experiments \cite{co98}. 
It is however not obvious that the good agreement with the FHNC results
obtained for the ground state can be conserved also for the excited states. 
Since complete FHNC calculations are available for one-particle one-hole
(1p-1h) nuclear matter charge responses \cite{fan87} we have
applied our model to calculate these responses.

The basic idea of the model, already presented in \cite{co95} and
\cite{co98}, consists in truncating the CBF expansion in order to consider
only those terms containing a single Jastrow-type correlation line, 
$h(r)=f^2(r)-1$. In Fig.~\ref{fig1}
we show the diagrams we have retained in the present calculation. 
It is worth pointing out that by setting in each diagram the particle line
equal to the hole line, we reproduce the set of diagrams used to
calculate the ground state  properties of the charge operator
\cite{co95}. An important property of the expansion in powers of $h(r)$ 
is that the normalizations of the wave function are exactly preserved 
at each order \cite{valencia97}. In Ref. \cite{co95}, the nuclear charge 
was conserved as well as the proper normalization of the correlated 
many-body wave functions in the present calculation. 
On the contrary, the nuclear charge is not conserved in
Refs.~\cite{giu91,ryc95}, where the expansion adopted, and truncated
at the second order, is based on {\it the number of particles} of the cluster
and not on the powers of $h(r)$.

The FHNC and the model calculations of the charge responses have been done 
using the same correlation, i.e. the scalar part of a complicated 
state dependent correlation fixed to minimize the nuclear 
binding energy in a FHNC calculation with the Urbana V14 
nucleon--nucleon potential \cite{wir88}.

In Fig.~\ref{fig2} we compare the results of our model
with those obtained with a FHNC calculations. In this figure we
show the proton structure functions, therefore no electromagnetic
nucleon form factors have been included. The structure functions
have been calculated for three values of the momentum
transfer, and for Fermi momentum of $1.09$ fm$^{-1}$. We found this value
of the Fermi momentum adequate to describe the quasi--elastic responses of
$^{12}C$  \cite{ama94}. In the panel $(a)$ of Fig.~\ref{fig2}
the full lines represent the results of the model and the
dashed lines those of the FHNC calculation. The difference between
the two calculations are very small, and they are explicitly shown,
multiplied by a factor $10^5$, in the panel $(b)$.

In Fig.~\ref{fig3} we show the response functions calculated using
the electromagnetic nucleon form factors of Ref.~\cite{hoh76}.
These response functions have been obtained considering, in
addition to the proton structure functions shown in Fig.~\ref{fig2},
also the neutron contribution, which in any case, turns out to be
negligible in the longitudinal response.
The dashed lines show the Fermi gas responses, corresponding in
our model to the first diagram of Fig.~\ref{fig1}. The
dashed--dotted lines have been obtained adding the contribution of the
two-point diagrams, i.e. the diagrams multiplied by the factor
$1/2$ in Fig.~\ref{fig1}. The full lines have been obtained by
including all the diagrams of Fig.~\ref{fig1}.
The contribution of the two--point diagrams is partially cancelled
by the inclusion of the three--point diagrams. This is an effect
similar to that obtained in the calculation of the ground state
charge and momentum distributions \cite{co95}.

The results we have presented
show that a model considering only those terms with a single 
correlation line can reproduce extremely
well the FHNC charge response functions. This conclusion is consistent  
with the finding of Ref.~\cite{co95} for the ground state charge 
distribution. We should remark that our calculations have been done for 
the charge operator only. An extension of the calculation to evaluate 
responses induced by
other electromagnetic operators is necessary to test the validity
and the range of applicability of the model.

We like to stress again that a good agreement with the FHNC results 
as been obtained, most probably,  because, 
in our model, the proper normalization of the many--body wave function 
has been conserved by evaluating both two-- and three--point diagrams. 
Calculations of 1p--1h responses which include the two--point diagrams only, 
overestimate the effect of the correlations. One may expect that the same
problem could affect also the two--nucleon emission calculations, like those of
Refs.~\cite{giu91,ryc95}, where only two--point diagrams are considered.
On the other hand, there are indications that in the 2p-2h responses two-- and
three--point diagrams act differently than in the 1p-1h responses 
\cite{co98}.
A similar analysis, as the one performed here, for the 2p-2h response 
is needed  to further clarify the situation.

\vskip 2. cm

\newpage

\noindent

\vskip 0.5 cm

\begin{figure}

\caption{ \label{fig1} }
Diagrams considered in our model. The dotted lines represent the
correlation function. The oriented lines represent
particle and hole wave functions. The black circle
indicates an integration point while the black square indicate the
integration point where the charge operator is acting. 
\end{figure}
\begin{figure}
\caption{ \label{fig2} }
In the panel (a) the 1p-1h nuclear matter proton structure functions
calculated with the  present model (full lines) are compared with
those obtained from the FHNC calculation of Ref.~\cite{fan87}. The
calculations have been done for $q=300$,$400$ and $550$ MeV/c and
$k_F=1.09$ fm$^{-1}$. In the panel (b) we show the differences, multipied
by $10^5$, between the structure functions obtained with the present model
and those obtained using FHNC. 
\end{figure} 
\begin{figure}
\caption { \label{fig3} }
Nuclear matter longitudinal responses for $q=300$, $400$ and $550$ 
MeV/c  and $k_F=1.09$ fm$^{-1}$ calculated with the proton and neutron form 
factors of ref. \cite{hoh76}. The dashed lines represent the Fermi
gas responses, the dashed--dotted lines have been obtained adding the
two--point diagrams, while the full lines show the results of the
complete calculations where all the diagrams of Fig.~\ref{fig1}
have been considered. 

\end{figure}

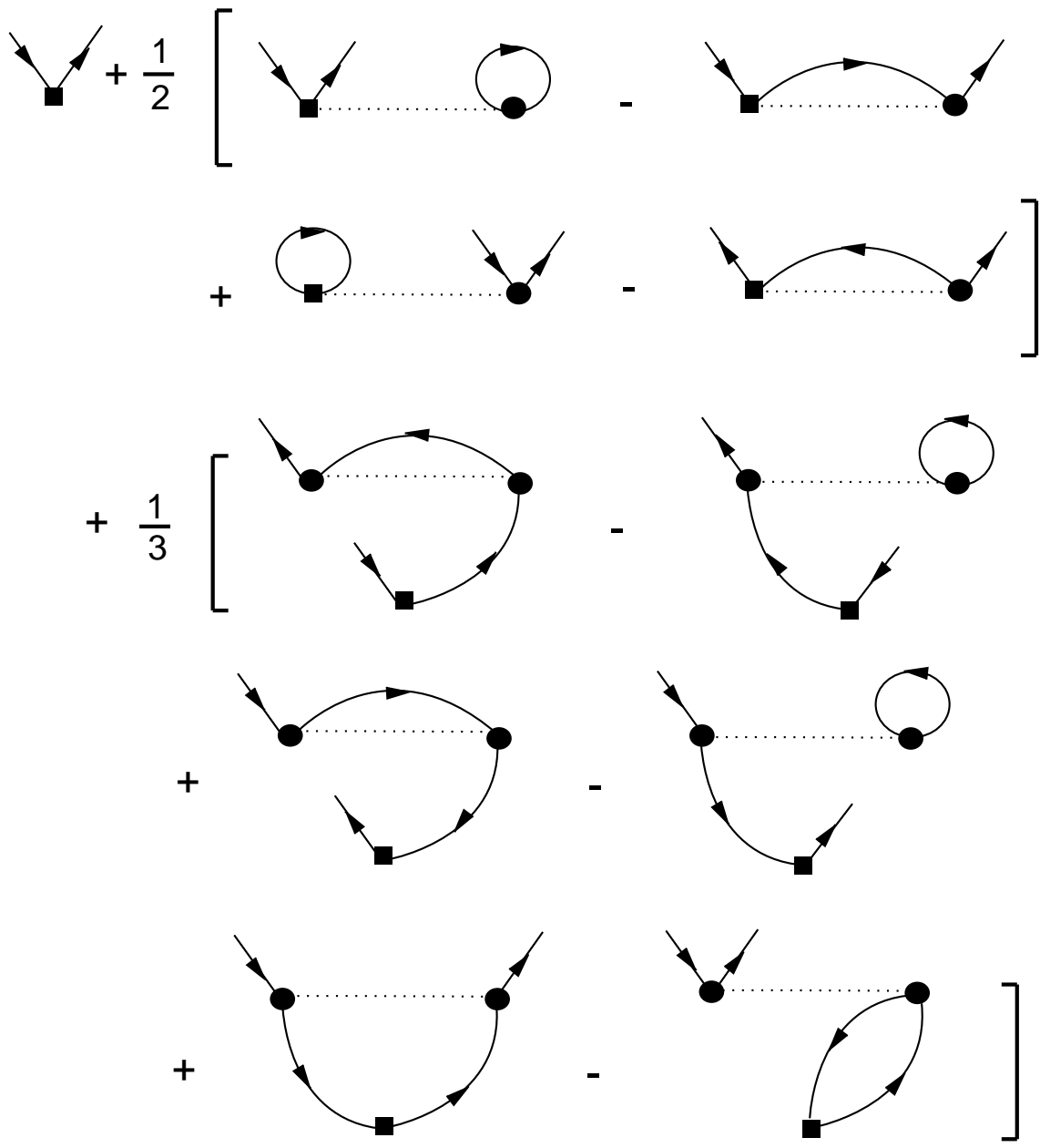
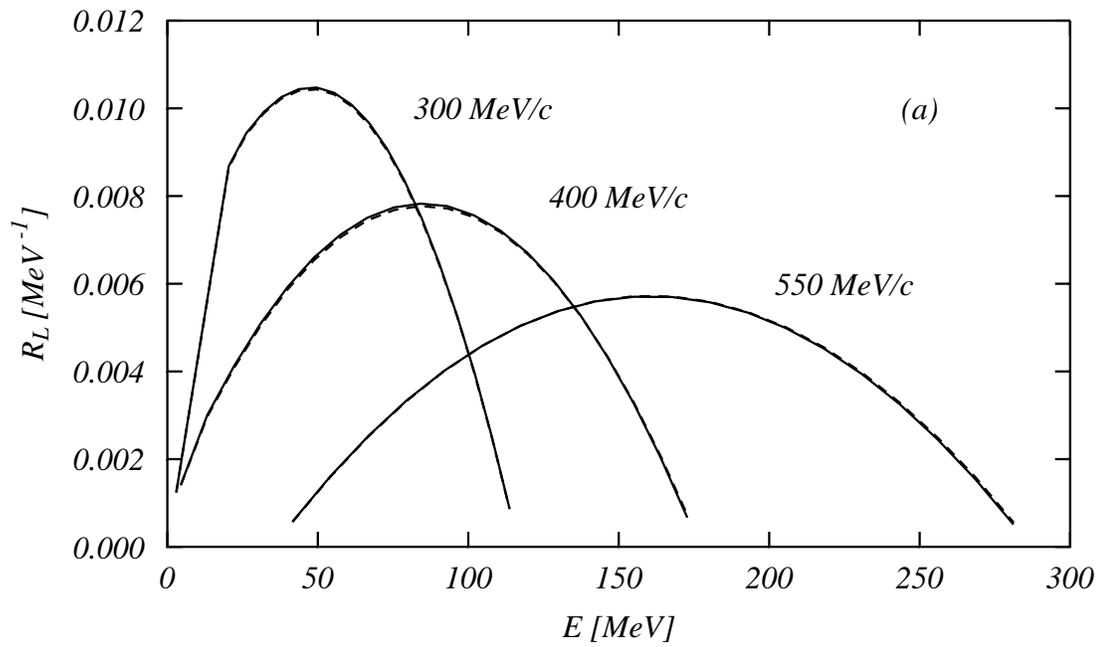
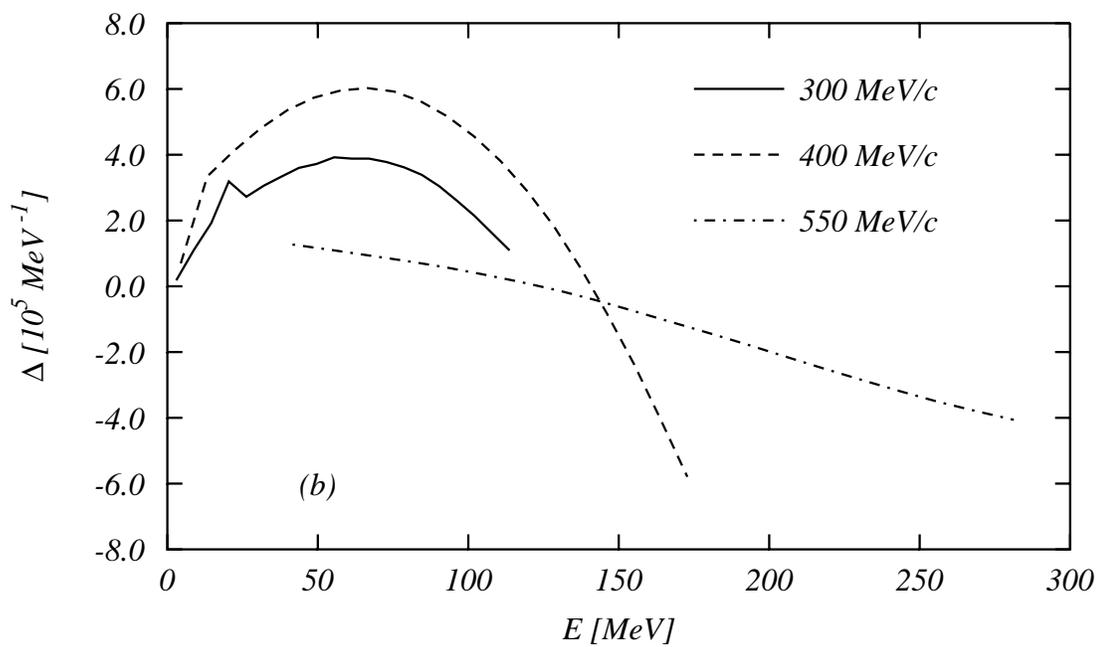
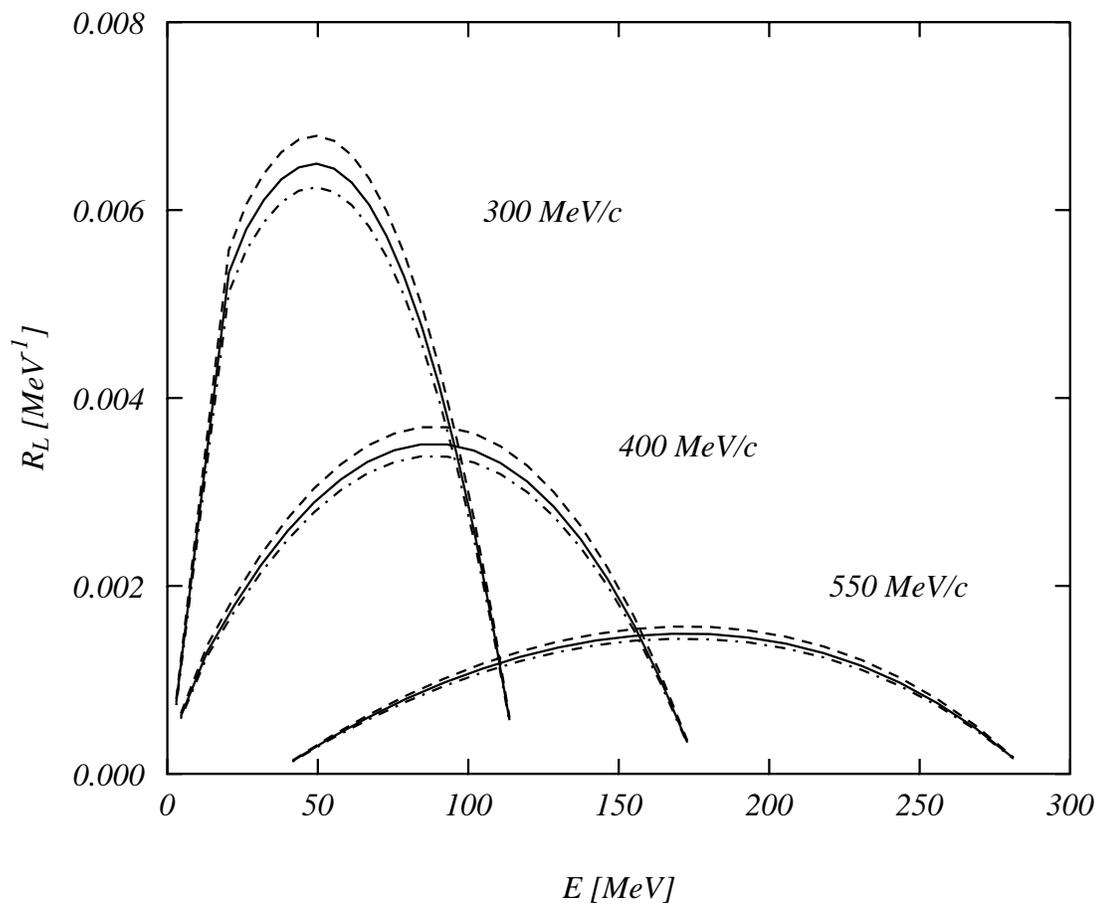
\end{document}